# Long Term H-Release from Amorphous Carbon Evidenced by in Situ Raman Microscopy under Isothermal Heating

C. Pardanaud[1], C. Martin[1], G. Giacometti[1], P. Roubin[1], B. Pégourié[2],

C. Hopf[3], T. Schwarz-Selinger[3], W. Jacob[3], J.G. Buijnsters[4]

[1] Aix-Marseille Université, CNRS, PIIM UMR 7345, 13397 Marseille cedex 20, France.

[2] CEA, IRFM, 13108 Saint-Paul-lez-Durance, France.

[3] Max-Planck-Institut für Plasmaphysik, EURATOM Association, Boltzmannstr. 2, 85748 Garching, Germany.

[4] Department MTM, Katholieke Universiteit Leuven, Kasteelpark Arenberg 44, B-3001 Leuven, Belgium.

**Abstract**

We study the kinetics of the H release from plasma-deposited hydrogenated amorphous carbon films under isothermal heating at 450, 500 and 600 °C for long times up to several days using in situ Raman microscopy. Four Raman parameters are analyzed. They allow the identification of different processes such as the carbon network reorganization and the H release from $sp^3$ or $sp^2$ carbon atoms and the corresponding timescales. Carbon reorganization with aromatization and loss of $sp^3$ hybridization occurs first in 100 minutes at 500 °C. The final organization is similar at all investigated temperatures. Full H release from $sp^3$ carbon occurs on a longer timescale of about 10 hours while H release from $sp^2$ carbon atoms is only partial, even after several days. All these processes occur more rapidly with higher initial H content, in agreement with what is known about the stability of these types of films. A quantitative analysis of these kinetics studies gives valuable information about the microscopic processes at the origin of the H release through the determination of activation energies.

## 1. Introduction

Raman spectroscopy is a powerful tool to rapidly characterize C-based materials that contain $sp^2$ carbon atoms ($C(sp^2)$). Interpreting the 1000 - 1800 $cm^{-1}$ spectral region gives information on the carbon organization (aromatization). Because the Raman cross section for $C(sp^3)$ is one to two orders of magnitude lower than that of $C(sp^2)$, two bands due to $C(sp^2)$ dominate this spectral range even when samples contain only a few percent of $C(sp^2)$. They are the well-known G and D bands [1]. They contain information on disorder such as the size of aromatic domains, the $C(sp^2)/C(sp^3)$ ratio for amorphous carbons (a-C), and the H content for hydrogenated carbons (a-C:H) [1-3]. Raman parameters generally used to probe the bonding structure are: the width and the position of the G and D bands, denoted $\Gamma_{G,D}$ and $\sigma_{G,D}$ respectively, the height ratio of these two bands, $H_D/H_G$, and if applicable, the $m/H_G$ parameter with m being the slope of the photoluminescence background calculated between 800 and 2000 $cm^{-1}$. In the case of heat-treated a-C:H films for which the H content varied from 2 to 30 at. %, we have recently obtained a linear relation between $\log_{10}(m/H_G)$ and the H content (H/H+C = 25 + 9 $\log_{10}(m/H_G)$) [4], similarly to what was previously obtained for as-deposited samples for which the H content varied from 20 to 47 at. % [5, 6]. We have emphasized that $m/H_G$ is sensitive not only to the H content, but also to defect passivation [7, 8], possibly by H migration, and we also found that $m/H_G$ is due only to H bonded to $C(sp^3)$. In addition we have obtained a linear relation between $H_D/H_G$ and the H content (H/H+C = 0.54 - 0.53 $H_D/H_G$), independent of whether the H is bonded to $C(sp^2)$ or to $C(sp^3)$. $H_D/H_G$ can thus be used to probe the total H content in the range from 2 to 30 at.%. Conversely, we have shown that the $\Gamma_G$ and the $\sigma_G$ parameters are sensitive mainly to the structural changes such as the re-organization leading to a larger aromatization and the loss of $C(sp^3)$ induced by heating.

To obtain more information on the different processes involved under heating we investigate here the long term kinetics of the isothermal hydrogen release from a-C:H films with initial H/H+C ratios of 0.29, 0.32 and 0.37 at temperatures of 450 °C, 500 and 600 °C by means of in situ Raman microscopy. We show how the evolution kinetics provide valuable information on the H release processes and emphasize the role of the initial structure and H-content on these processes.

## 2. Experimental

Plasma-deposited hydrogenated amorphous carbon films, a-C:H, with H/H+C ~ 0.29, 0.32 and 0.37 (DC self bias of –300, –200 and –100 V, respectively), and a thickness of e ~ 0.20 µm (see ref. [9] for details on the deposition method) were studied together with a thicker film with H/H+C = 0.32 (applied substrate bias : –200 V) and a thickness of e ~ 1 µm (see ref. [5] for details on the deposition method). The 0.29 and 0.32 films are hard films while the 0.37 film is intermediate between a hard and a soft film [9]. Note that the thick film has a larger $C(sp^3)$ content than the three thin films: this is indicated by its G band position at 1540

cm$^{-1}$ instead of 1522 cm$^{-1}$ for the thin films, their G band widths being similar (see figure 9 in [2]). The samples with the as-deposited films were cut into several pieces, each piece being heated under argon flow in a cell at 1.5 bar with a linear ramp of 150 °C min$^{-1}$ to reach the working temperature. Time zero was taken at the end of the ramp. Temperatures 450, 500 and 600 °C were studied.

Raman spectra were obtained using a Horiba-Jobin-Yvon HR LabRAM apparatus (laser wavelength: $\lambda_L$ = 514.5 nm, 50X objective, numerical aperture of 0.5 leading to a laser focus diameter of 2.5 µm, resolution ~ 1 cm$^{-1}$). The laser power was chosen to have a good signal/noise ratio and/or to prevent damages and three values were tested, 0.01, 0.1 and 1 mW, corresponding to power densities from $2 \times 10^2$ to $2 \times 10^4$ J cm$^{-2}$ s$^{-1}$. Unless otherwise stated, spectra were recorded in situ at the working temperature. The cell was a commercial LINKAM TS1500, argon was Alphagaz 2 (99.9995% purity, Air Liquid Company). The Raman parameters analyzed were the G band wavenumber, $\sigma_G$, the G band full-width at half-maximum, $\Gamma_G$, the relative height of the G and D bands, $H_D/H_G$, and the m/$H_G$ parameter. Heights were measured on the raw data after the linear baseline calculated between 800 and 2000 cm$^{-1}$ (slope m) was subtracted. In that case, $H_D$ was therefore measured at its apparent maximum, except when the D band maximum was not sufficiently well defined (i.e. for the as deposited sample). In the latter case $H_D$ was taken at 1370 cm$^{-1}$. All these parameters are presented in figure 1.

## 3. Results and discussion

Figure 1 displays Raman spectra of the a-C:H film (e = 0.2 µm, H/H+C = 0.32), as deposited at room temperature and after 4, 45 and 500 minutes at 500 °C. The Raman spectrum of the as deposited film is composed of the underlying silicon wafer signature at 520 cm$^{-1}$, a G band at 1522 cm$^{-1}$, and a broad shoulder at ~ 1200 – 1400 cm$^{-1}$ containing the D band. When the sample is heated, the silicon signature diminishes because the a-C:H becomes more absorbent [10, 11]. The photoluminescence background significantly increases, due to the defect passivation (and the quenching of non-radiative relaxation processes [4, 6]) before it starts decreasing as hydrogen is released. The G band narrows and slightly blueshifts, as previously observed in [4, 12]. The D band also narrows and its height increases. According to previous studies [2, 12, 13], these evolutions reveal the organization of the material, i.e. a loss of C(sp$^3$) and increasing aromatization. When the heating time increases, the Raman spectra keep changing: the G and D bands become more clearly distinguishable, $H_D/H_G$ increases and $m/H_G$ decreases. A long timescale process thus occurs, that we investigate in more detail below.

Figure 2 displays the isothermal evolution with time up to 800 min of $\sigma_G$, $\Gamma_G$, $m/H_G$ and $H_D/H_G$ at 500 °C at the three laser powers, 0.01, 0.1 and 1 mW for the a-C:H film (e = 0.2 µm, H/H+C = 0.32). No significant effect of the laser power is observed, indicating that no additional heating due to the laser irradiation is induced even at 1 mW, contrary to what was measured in [14] for thinner (15 nm) films. Figures 2.a and 2.b display a similar isothermal evolution for $\Gamma_G$ and $\sigma_G$, reaching plateaus in ~ 100 minutes. $\Gamma_G$ is at 180 cm$^{-1}$ for the as-deposited film and the plateau at 500 °C is at 85 cm$^{-1}$. $\sigma_G$ is at 1522 cm$^{-1}$ for the as-deposited film and the plateau at 500 °C is at 1580 cm$^{-1}$. These evolutions suggest that at 500 °C, the film gets rapidly more organized (in typically 100 min) and that this organization then does not significantly evolve. Conversely, figures 2.c and 2.d show that for $m/H_G$ and $H_D/H_G$ this rapid evolution is followed by a slow and roughly linear change (decrease for $m/H_G$ and increase for $H_D/H_G$). $H_D/H_G$ and $m/H_G$ are both correlated with the H content [4]. We thus interpret these evolutions as due to the isothermal H release, with a first phase occurring rapidly together with structural changes (within about 100 minutes), followed by a second phase occurring on a longer time scale.

To test the spatial homogeneity of heating-induced changes, we map the film at room temperature after heating at 500 °C for 800 min, in a ~ 35 x 35 µm$^2$ area, much larger than the 5 µm$^2$ area of the laser spot, with the 0.01 mW laser power. Histograms obtained by analyzing the maps of the four parameters, $\sigma_G$, $\Gamma_G$, $m/H_G$ and $H_D/H_G$ are displayed in figure 3. Their widths are small ($\sigma_G$ = 1601 ± 2 cm$^{-1}$, $\Gamma_G$ = 76 ± 4 cm$^{-1}$, $m/H_G$ = 0.25 ± 0.09 µm and $H_D/H_G$ = 0.77 ± 0.02), indicating that the sample is uniform, both in structure and in composition. The values obtained for $\Gamma_G$, $m/H_G$ and $H_D/H_G$ at the end of the 800 min heating at 500 °C fall in this range, confirming that the spot irradiated by the laser is not more heated than the other parts of the sample. Conversely, $\sigma_G$ is at 1580 cm$^{-1}$ at the end of the heating process, significantly outside the histogram values, and this is due to a reversible temperature effect due to multi-phonon interactions, similarly to what was found for graphite or graphene [15]. We have checked that the effect was reversible by varying alternately the temperature between room temperature and 500 °C.

Figure 4 is a plot similar to that of figure 2 to compare the isothermal evolutions of the three thin films (e = 0.2 µm and H/H+C = 0.29, 0.32 and 0.37) at 500 °C and that of the thick film (e = 1 µm and H/H+C =

0.32) at 450 °C. The laser power was 1 mW, except for the first 400 min at H/H+C = 0.29 for which it was only 0.1 mW (as the spot was located at another place on the sample, the small jump in the data is due to the small width of the structure and H-content distribution, see figure 3). For all the Raman parameters at 500 °C and for the first 100 s the evolution is the faster the higher the initial H content. This is in agreement with what is known about the stability and the H desorption of this type of hydrogenated films [16, 17], the higher the H content, the less stable is the film and the more rapid is the H desorption. For longer times, the plateaus reached by $\sigma_G$ and $\Gamma_G$ are almost independent of the initial H content, indicating that the structure of the film at 500 °C is independent of the initial H content. $m/H_G$ goes to zero for the three films at 500 °C while $H_D/H_G$ keeps slowly increasing along three distinct lines. As $H_D/H_G$ depends on to the total H content and $\log_{10}(m/H_G)$ only on H bonded to $C(sp^3)$, this indicates that at this stage of the kinetics all remaining H is bonded to $C(sp^3)$. Furthermore, the remaining H content is higher for lower initial H content. The thick film evolution of $\sigma_G$ and $\Gamma_G$ and $m/H_G$ at 450 °C is intermediate between the thin films having the H content 0.29 and 0.32 at 500 °C, although the first stage of the kinetics during the first min is faster in the case of the $m/H_G$ and $H_D/H_G$ parameters. The $H_D/H_G$ parameter also indicates that the H becomes significantly lower. This film has a slightly different carbon structure ($C(sp^3)$ / $C(sp^2)$ ratio as indicated by the value of the G band position, see experimental section) than the other films but also a different thickness. Note that this type of thick film is known to easily delaminate, which might also play a role on the H-release, this analysis being beyond the scope of this paper. $\sigma_G$ and $\Gamma_G$ are known to give similar information [18] and we will use only $\Gamma_G$ in what follows.

We now focus on the long term evolutions at different temperatures. Figure 5 displays the isothermal evolution with time up to 6000 minutes of $\Gamma_G$, $m/H_G$ and $H_D/H_G$ at 450, 500 and 600°C for the a-C:H film with e = 0.2 µm, H/H+C = 0.29. As expected, the higher the temperature, the faster the kinetics. A first order exponential law correctly fits the data giving the following estimation of characteristic times: 1370, 140, 2 min for $\Gamma_G$, 3340, 190, 4 min for $\log(m/H_G)$, and 2350, 730, 65 min for $H_D/H_G$, at 450, 500 and 600 °C, respectively. The same plateau is found for the three temperatures for $\Gamma_G$. This is also the case for $m/H_G$ although the effect is less clear because the signal to noise ratio is significantly lower for the 500 °C data than for the other temperatures. Conversely, the asymptotic value of $H_D/H_G$ increases when the temperature increases. The behaviour of $\Gamma_G$ suggests that the final stage of organization is the same for all investigated temperatures. More precisely, as $\Gamma_G$ (as well as $\sigma_G$) is related to $C(sp^3)$ defects linked to the aromatic network, it suggests that the vanishing of $C(sp^3)$ defects and the aromatization occur similarly in this temperature range. $m/H_G$ is also related to $C(sp^3)$, probing H coming only from H bonded to $C(sp^3)$ atoms and consistently, it behaves as $\Gamma_G$, going to zero at all temperatures. $H_D/H_G$ probes the full H content, and its different final values suggest that not all the H bonded to $C(sp^2)$ can be released in this temperature range, even with infinite time. Applying the relation between the H content and $H_D/H_G$ (see [4]), the H contents can be estimated at 12 and 8 % at 500 °C and 600 °C, respectively. At 450°C, 6000 minutes are not enough to reach a plateau.

Even though the understanding of the characteristic times obtained in figure 5 is not straightforward, these times may give information on the processes driving the H release. We have plotted in figure 6 the logarithm of the inverse of these times as a function of the inverse of temperature. The linearity observed indicates that typical activation energies could be deduced, here 2.4 eV for $\Gamma_G$ and $\log_{10}(m/H_G)$ and 1.3 eV for $H_D/H_G$. A detailed analysis of these observations needs additional data and is beyond the aim of this paper.

## 4. Conclusion

The isothermal evolutions of the bonding structure and the H content of a-C:H films with various H contents (H/H+C = 0.28, 0.32 and 0.37) and various thicknesses (e = 0.2 µm and 1 µm) were studied in the range 450 °C – 600 °C by means of Raman microscopy. At 500 °C, a fast process occurring in less than 100 minutes is evidenced. This is attributed to reorganization of the carbon network, i.e. loss of $C(sp^3)$ and increasing aromatization. During this reorganization, H begins to be released, the full H release from $C(sp^3)$ occurring in ~ 15 hours. On longer timescale (~ 50 hours), H keeps being released, but ~ 12 % of H bonded to $C(sp^2)$ finally remain. Kinetics is similar, but more (less) rapid, at 600 (450) °C. Kinetics is also faster when the H content is higher, in agreement with what is known on the instability of this type of hydrogenated films when the H content is large. Not only the carbon network organization but also the initial $C(sp^3)/C(sp^2)$ ratio can significantly changes the H release kinetics. This type of isothermal kinetics studies reveal information on the processes involved in the H release or in the carbon organization under heating, for example through the determination of activation energies. Additional experiments are currently performed at other temperatures and also with deuterated films to give more data to elucidate these thermally activated processes.

**Acknowledgments**

We acknowledge the Euratom-CEA association, the EFDA European Task Force on Plasma Wall Interactions, the Fédération de Recherche FR-FCM, the French agency ANR (ANR-06-BLAN-0008 contract) and the PACA Region (FORMICAT project) for financial support. JGB acknowledges the Executive Research Agency of the European Union for funding under the Marie Curie IEF grant number 272448.

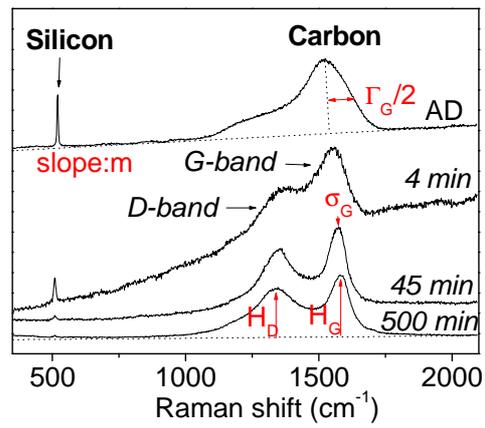

**Figure 1.** Raman spectra of the a-C:H film (e = 0.2 μm and H/H+C = 0.32) at room temperature, as deposited (AD) and at 500 °C after 4, 45 and 500 min at 500 °C.

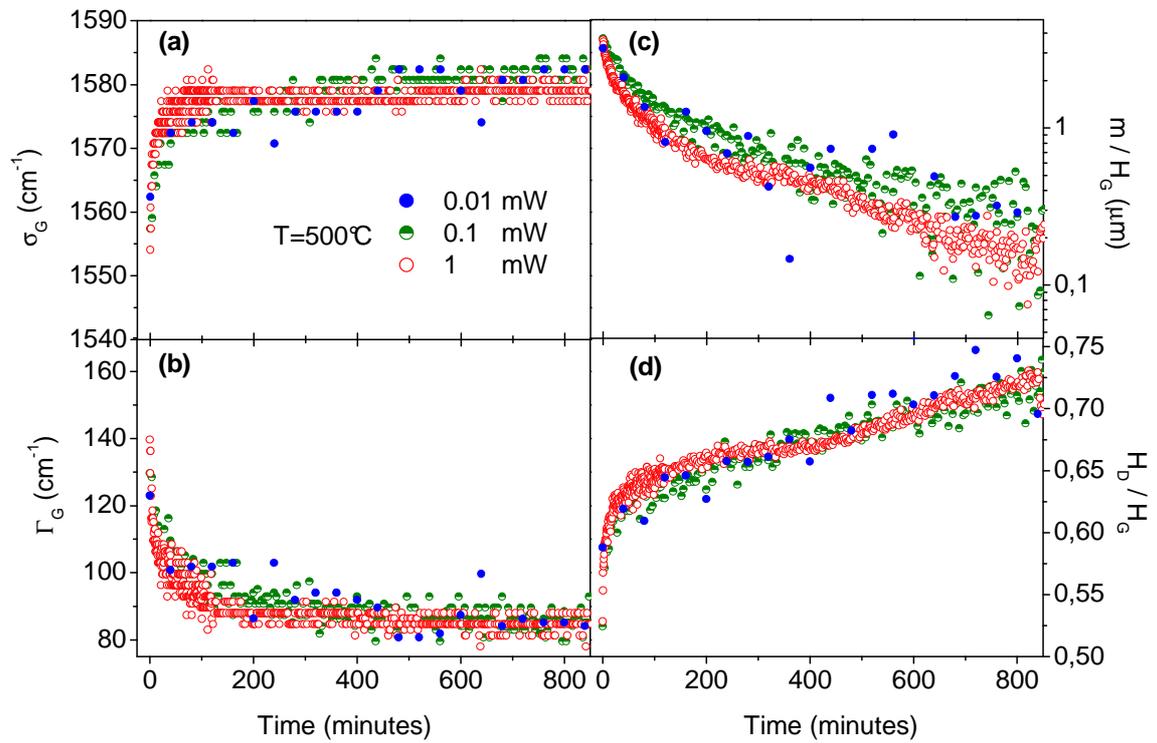

**Figure 2.** Time evolution of Raman parameters of the a-C:H film (e = 0.2 μm and H/H+C = 0.32) heated at 500 °C at three laser powers: (a) G band position, $\sigma_G$, (b) G band width, $\Gamma_G$, (c) $m/H_G$ parameter, m being the slope of the photoluminescent background and (d) D and G band height ratio, $H_D/H_G$.

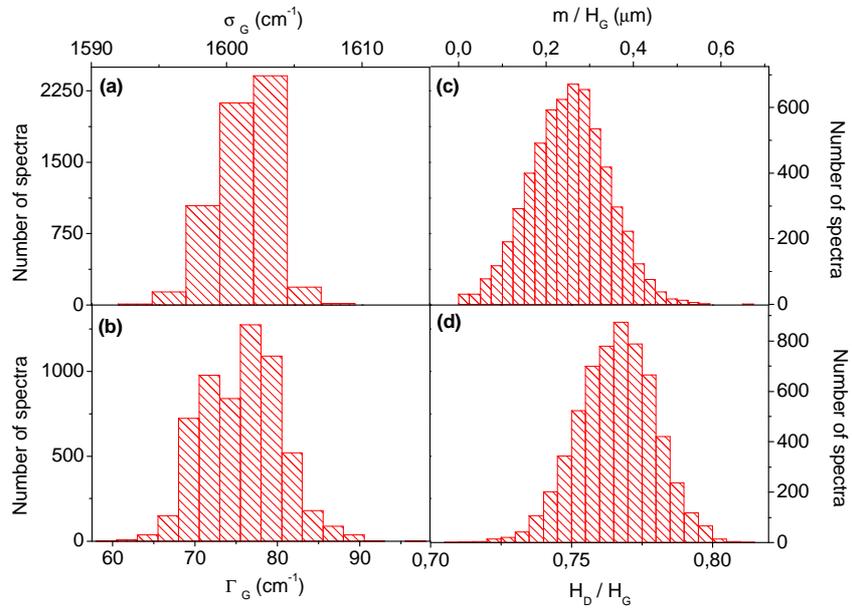

**Figure 3.** Histograms of the mapping of Raman parameters for the a-C:H film of figure 2 at room temperature after 800 minutes at 500°C. Spectra were recorded on a 35 x 35 μm$^2$ area, with a 0.5 μm lateral sampling and a laser power of 0.01 mW. (a) G band position, $\sigma_G$, (b) G band width, $\Gamma_G$, (c) m/H$_G$ parameter, m being the slope of the photoluminescent background and (d) D and G band height ratio, H$_D$/H$_G$.

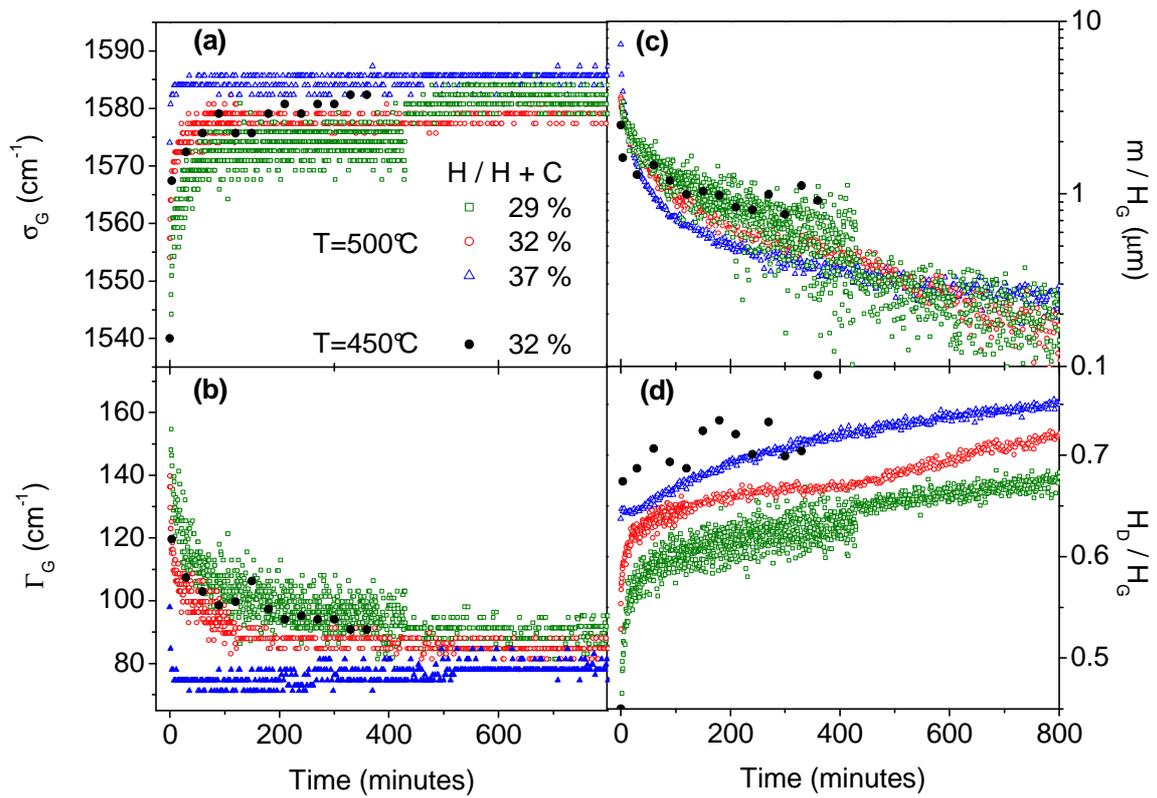

**Figure 4.** Same as figure 2 for the three thin films (e = 0.2 mm and H/H+C = 0.29, 0.32 and 0.37) at 500 °C and for the thick film (e = 1 mm and H/H+C = 0.32) at 450 °C.

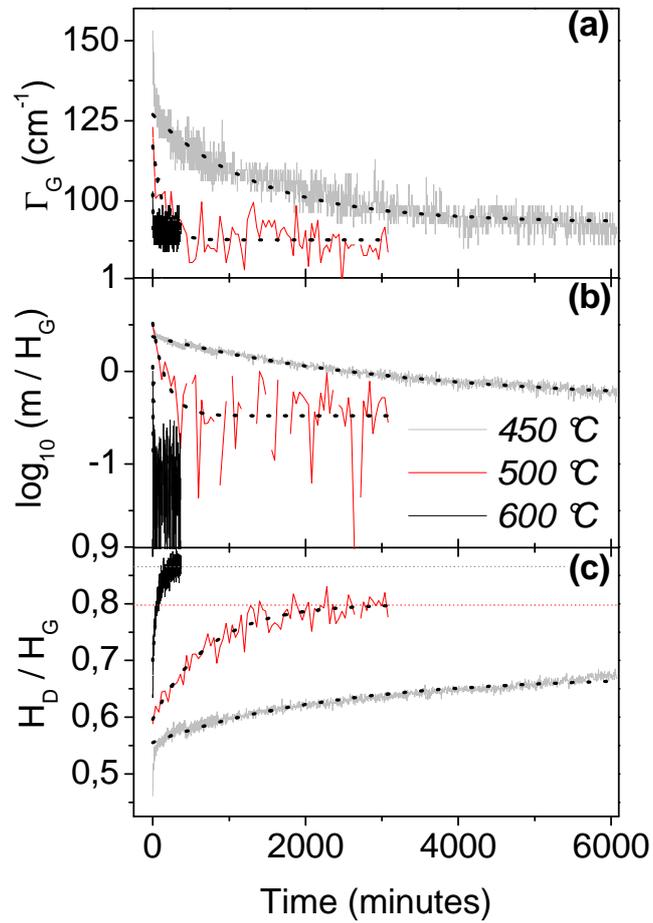

**Figure 5.** Time evolution of Raman parameters of the a-C:H film (e = 0.2 μm and H/H+C = 0.29) heated at 450, 500 and 600 °C: (a) G band width, $\Gamma_G$, (b) $m/H_G$ parameter and (c) D and G band height ratio, $H_D/H_G$. Black dotted lines are exponential fits.

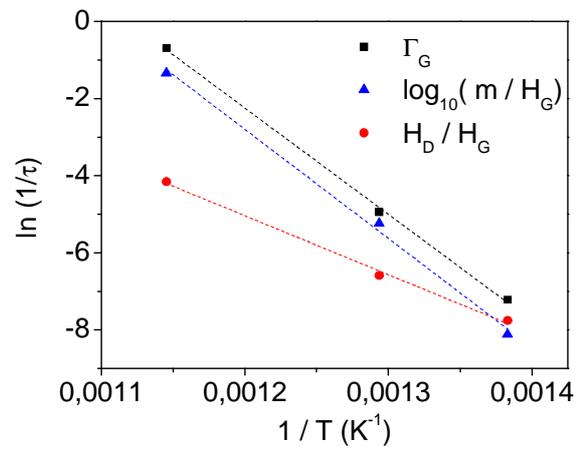

**Figure 6**. Arrhenius plot of data from the exponential fits of figure 5.